\documentclass[prl,twocolumn,preprintnumbers,amsmath,amssymb,showpacs,superscriptaddress]{revtex4} 
\usepackage{epsfig}
\usepackage{amssymb}
\usepackage{amsmath}
\usepackage{bm}
\usepackage{color}
\usepackage{times}
\newcommand{\be}{\begin{equation}}
\newcommand{\ee}{\end{equation}}
\newcommand{\bea}{\begin{eqnarray}}
\newcommand{\eea}{\end{eqnarray}}
\newcommand{\up}{\uparrow}
\newcommand{\down}{\downarrow}

\def\nn{\nonumber\\}
\def\fr#1{(\ref{#1})}

\def\kf{k_{\rm F}}
\begin{document}
%%%%%%%%%%%%%%%%%%%%%%%%%%%%%%%%%%%%%%%%%%%%%%%%%%%%%%%%%%%%%%%%%%%%%%%%%%%%%%
\title{Shell-Filling Effect in the Entanglement Entropies of Spinful Fermions} 
%%%%%%%%%%%%%%%%%%%%%%%%%%%%%%%%%%%%%%%%%%%%%%%%%%%%%%%%%%%%%%%%%%%%%%%%%%%%%%
\author{Fabian H.L. Essler}
\affiliation{The Rudolf Peierls Centre for Theoretical Physics, Oxford
University, Oxford OX1 3NP, UK}
\author{Andreas M. L\"auchli}
\affiliation{Institut f\"ur Theoretische Physik, Universit\"at Innsbruck, A-6020 Innsbruck, Austria}
\author{Pasquale Calabrese}
\affiliation{Dipartimento di Fisica dell'Universit\`a di Pisa and INFN, 56127 Pisa, Italy}
\begin{abstract}
We consider the von Neumann and R\'enyi entropies of the one
dimensional quarter-filled Hubbard model. We observe that for periodic
boundary conditions the entropies exhibit an unexpected dependence on
system size: for $L=4\ {\rm mod\ 8}$ the results are in agreement with
expectations based on conformal field theory, while for $L=0\ {\rm
  mod\ 8}$ additional contributions arise. We explain this observation
in terms of a shell-filling effect, and develop a conformal field theory
approach to calculate the extra term in the entropies. Similar shell filling effects
in entanglement entropies are expected to be present  in higher dimensions
and for other multicomponent systems.
\end{abstract}
\maketitle

Over the course of the last decade, entanglement measures have
developed into a powerful tool for analyzing many-particle quantum
systems, in particular in relation to quantum criticality and
topological order \cite{rev}. Within the realm of one dimensional (1D)
systems, arguably the most 
important result concerns the universal behaviour in critical
theories, which is characterized by the central charge of the
underlying conformal field theory (CFT) \cite{CFT0,CFT,CFT2}. Let
consider the ground state $|{\rm GS}\rangle$ of a finite, periodic 1D
system of length $L$ and partition the latter into a finite block $A$
of length $\ell$ and its complement $\bar{A}$. The density matrix of
the entire system is then $\rho=|{\rm GS}\rangle\langle{\rm GS}|$, and
we will denote the reduced density matrix of block $A$ by
$\rho_A$. Widely used measures of entanglement are the R\'enyi
entropies 
\be
S_n = \frac{1}{1-n}\ln[{\rm Tr}{\rho_A^n}]\ .
\ee
They encode the full information on the spectrum of $\rho_A$
\cite{cl-08}, and in the limit $n\to 1$ reduce to the von Neumann
entropy $S_1=-{\rm Tr}{\rho_A\ln\rho_A}$. When the subsystem size
$\ell$ is large compared to the lattice spacing, $S_n$ are given by
\be
S_{n} =\frac{c}{6}\left(1+\frac{1}{n}\right) \ln \Big(\frac{L}\pi
\sin\frac{\pi \ell}L\Big) +c'_n+o\big(1\big)\,, 
\label{criticalent}
\ee
where $c$ is the central charge, $c'_n$ are non-universal additive
constants, and $o(1)$ denotes terms that vanish for $\ell\to\infty$. The
result \fr{criticalent} has been confirmed for many spin-chains and
itinerant lattice models, see \cite{rev} for recent reviews. The
knowledge of the entanglement entropies has led to a deeper
understanding of numerical algorithms based on matrix product states
\cite{mps} and has aided the development of novel computational
methods \cite{cirac}. 

The Hubbard model is a central paradigm of strongly correlated
electron systems. Its 1D version has attracted much attention for
decades, because it is exactly solvable and exhibits a Mott metal to
insulator transition \cite{book}. The Hamiltonian for periodic
boundary conditions is
\be
H_{\rm Hubb}=-t\sum_{j=1}^L\sum_{\sigma}c^\dagger_{j,\sigma}c_{j+1,\sigma}+{\rm h.c.}
%c^\dagger_{j+1,\sigma}c_{j,\sigma}
%\nn&&
+U\sum_j n_{j,\uparrow}\ n_{j,\downarrow},%-\mu\sum_j n_j\ ,
\label{HHubb}
\ee
where $c^\dagger_{j,\sigma}$ are fermionic spin-$\frac{1}{2}$ 
creation operators at site $j$ with spin $\sigma=\up,\down$,
$n_{j,\sigma}=c^\dagger_{j,\sigma}c_{j,\sigma}$, and we will assume
repulsive interactions $U\geq 0$. In the following we will for the
sake of definiteness fix the band filling to be one electron per two
sites, i.e. $N_\uparrow=N_\downarrow=\frac{L}{4}$, but we stress that
our findings generalize to other fillings and, in fact, to other
models. It is known from
the exact solution that the ground state of \fr{HHubb} below half
filling (less than one fermion per site) is metallic and the low
energy physics of the model is described by a spin and charge
separated Luttinger liquid \cite{book} equivalent to the semi-direct
product of two $c=1$ CFTs \cite{FSspectrum}.

\begin{figure}[b]
\begin{center}
%\epsfxsize=0.42\textwidth
%\epsfbox{Fig1}
\includegraphics[width=0.95\linewidth]{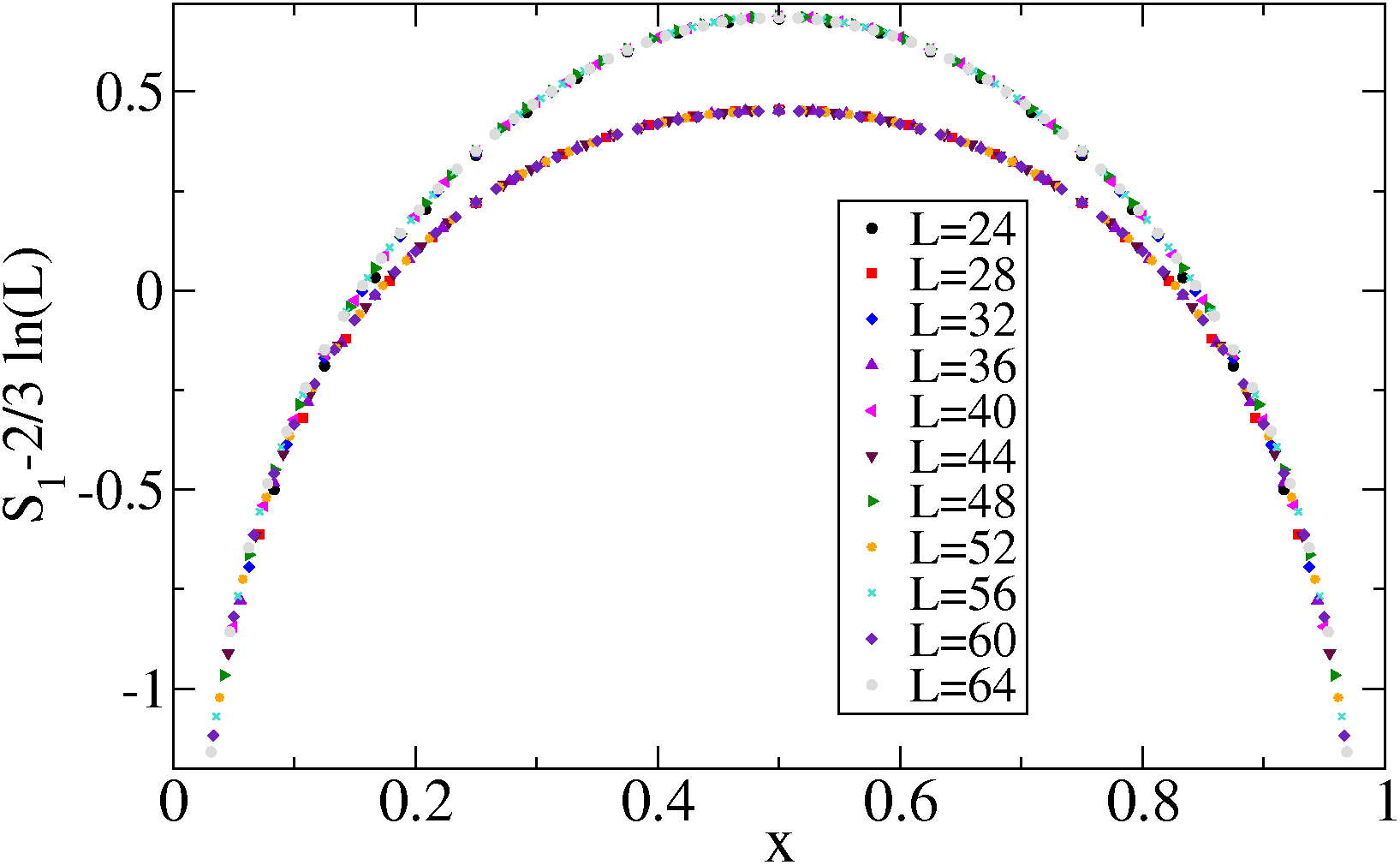}
\end{center}
\caption{DMRG data for $S_1-2/3 \ln L$ as a function of $x=\ell/L$ for $U=t$ and
$L=24,28,32,36,40,44,48,52,56,60,64$. The lower and upper branches
corresponds to lattice lengths $L=4\ {\rm mod}\ 8$ and $L=0\ {\rm
  mod}\ 8$ respectively.} 
\label{fig:vN1s}
\end{figure}

Given this state of affairs, it is quite surprising that the
entanglement entropies do not always follow (\ref{criticalent}). This
is shown in Fig.~\ref{fig:vN1s}, which shows 
numerical results for $S_1$ obtained by density matrix renormalization
group (DMRG) for a quarter-filled Hubbard model at $U=t$ for a number
of different lattice lengths $L$. The data is seen to collapse on
\emph{two} curves, with only the lower one following the CFT result
(\ref{criticalent}).  
Interestingly, both the $L=4\ {\rm mod}\ 8$ and the $L=0\ {\rm
  mod}\ 8$ data exhibit scaling collapse, but to different functions.
 We stress that this behaviour is very different from the lattice
``parity effects''  for Luttinger liquids %reported in Refs.~
\cite{oscillations,osc2}, which refer to $o(1)$ corrections in
$S_{n\geq 2}$ only. 

{\it The shell-filling effect}.
In order to understand the origin of the difference in entanglement
entropies between $L=4\ {\rm mod}\ 8$ and $L=0\ {\rm mod}\ 8$, we
consider the ground state in the limit $U\to 0$. Here we are dealing
with with non-interacting, spinful fermions, for which the boundary
conditions on a ring fix the momenta to be $p_m=2\pi m/L$ with integer
$m\in [-L/2,L/2)$. For a chain of length $L=8n+4$, quarter filling
corresponds to an odd number $N_\sigma=L/4=2n+1$ of spin-$\sigma$
fermion, and the unique ground state is the {\it symmetric} Fermi sea    
\be
|2n+1\rangle_{\rm FS}= \prod_{m=-n}^n
c^\dagger_\uparrow(p_m)c^\dagger_\downarrow(p_m)|0\rangle\ ,
\label{gsbad}
\ee
where $c^\dagger_\sigma(k)=L^{-1/2}\sum_{j=1}^L e^{-ikl}c^\dagger_{j,\sigma}$ are
creation operators in momentum space and $|0\rangle$ is the fermionic
vacuum state. On the other hand, when $L=8n$, $N_\sigma=L/4=2n$ is
even and it is impossible for a given spin species to form a symmetric
Fermi sea. As a result the ground state is degenerate. In particular,
there are two degenerate ground states with $N_\sigma=L/4=2n$, that
have zero momentum and are parity eigenstates (parity is a good
quantum number) 
\bea
|\sigma\rangle=\frac{c^\dagger_\up(\kf)c^\dagger_\down(-\kf)+\sigma
  c^\dagger_\down(\kf)c^\dagger_\up(-\kf)}{\sqrt{2}}
|2n-1\rangle_{\rm FS}.
\eea
Here $k_F=\pi/4$ is the Fermi momentum. As is shown below, the
$U\to 0$ limit of the Hubbard model ground state gives the state
$|+\rangle$. The shell-filling effect is now clear: for 
$L=4\ {\rm mod}\ 8$ the ground state is a symmetrically filled Fermi
sea, while for $L=0\ {\rm mod}\ 8$ it is given by the linear
superposition of two asymmetric Fermi seas. In terms of spin symmetries 
this state corresponds to the $S^z=0$ component of a $S=1$ multiplet.

{\it Bethe Ansatz (BA) solution}. We now turn to the case
$U>0$. Eigenstates of the Hubbard chain are parametrized in terms of
the solutions $\{\Lambda_\alpha,k_j \}$ of the following set of
coupled BA equations \cite{lw,book}
\begin{eqnarray} 
&&     k_j L  =  2 \pi I_j - \sum_{\alpha = 1}^{N_\down}
                 \theta \Big(
		 \frac{\sin k_j - \Lambda_\alpha}{u} \Big),\quad
               j=1,\ldots,N\ ,
		 \nn
&&     \sum_{j=1}^{N} \theta \Big(
		 \frac{\Lambda_\alpha - \sin k_j}{u} \Big)  = 
		 2 \pi 
		 J_\alpha+
		 \sum_{\beta = 1}^{N_\down}
		 \theta \Big(
		 \frac{\Lambda_\alpha - \Lambda_\beta}{2u} \Big) ,\nn
&&\hskip 4cm\alpha=1,\ldots,N_\down.
\label{BAE}
\end{eqnarray}
Here 
$u={U}/({4t})$, $\theta(x) =2 \arctan(x)$ and $N=N_\up+N_\down$. For
real solutions of the BA equations \fr{BAE}, the
``quantum numbers'' $I_j$ ($J_\alpha$) are integers if $N_\down$ is even (if
$N_\up$ is odd) and half-odd integers if $N_\down$ is odd (if $N_\up$
is even). The momentum is expressed in terms of the parameters
$\{\Lambda_\alpha,k_j \}$ by $P=\sum_{j=1}^N k_j$, while the energy
(in units of $t$) is given by
\be
E=uL-\sum_{j=1}^N\left[2\cos(k_j)+\mu+2u\right]\ ,
\label{EP}
\ee
where $\mu$ is the chemical potential. Following Ref. \cite{hwt}, we
define \emph{regular} BA states as eigenstates of Eq. (\ref{HHubb})
arising from solutions of \fr{BAE} with $2N_\down\leq N$, where all
$k_j$ and $\Lambda_\alpha$ are finite. We denote these states by
$|\{I_j\};\{J_\alpha\}\rangle_{\rm reg}$. As was shown in
Ref.~\cite{hwt}, all regular BA states are lowest-weight states with
respect to the SO(4) symmetry of the Hubbard model \cite{so4}, and a
complete set of energy eigenstates is obtained by acting on them with
the SO(4) raising operators. For $L=4\ {\rm mod}\ 8$ it
is known \cite{lw,book} that the quarter filled ground state is a
regular BA state characterized by the choice
$I_j=-2n-\frac{3}{2}+j\ ,\ j=1,\ldots,4n+2$ and
$J_\alpha=-n-1+\alpha\ ,\ \alpha=1,\ldots,2n+1$.  

For $L=8n$ ($n$ a positive integer), we find that there are two
degenerate lowest energy regular, real solutions of \fr{BAE}
with $N_\up=N_\down=2n$ fermions. They are obtained by the two choices
$J_\alpha^{(1,2)}=-n-\frac{1}{2}+\alpha\ ,\ \alpha=1,\ldots,2n$ and 
$I_j^{(1)}=-2n+j\ ,\ j=1,\ldots,4n$ or  
$I_j^{(2)}=-2n-1+j\ ,\ j=1,\ldots,4n$.
We stress that the distribution of the $I_j$ is asymmetric around
zero in both cases. Interestingly, these are not ground states. The
regular solution with the lowest energy involves one pair of complex
conjugate $\Lambda_\alpha$'s known as a 2-string, but it is not the
ground state either. 

Let us now consider regular BA states with total spin quantum number
$S^z=1$, i.e. $N_\up=2n+1$, $N_\down=2n-1$. These are by construction
lowest weight states of the spin-SU(2) symmetry algebra.
The lowest energy regular BA state in this sector corresponds to the
(symmetric) choice 
$I_j^{(0)}=-2n-\frac{1}{2}+j\ ,\ j=1,\ldots,4n$, and 
$J^{(0)}_\alpha=-n+\alpha\ ,\ \alpha=1,\ldots,2n-1$.
Crucially, the state
\be
S^-|\{I_j^{(0)}\};\{J_\alpha^{(0)}\}\rangle_{\rm reg},
\label{gsH}
\ee
is a (non-regular) eigenstate of the Hubbard Hamiltonian with
$N_\up=N_\down=L/8$ fermions. Here
$S^-=\sum_{j=1}^Lc^\dagger_{j,\down} c_{j,\up}$ is the spin lowering operator.
As $[S^-,H]=0$ its energy is the same as
that of the regular BA state $|\{I_j^{(0)}\};\{J_\alpha^{(0)}\}\rangle_{\rm reg}$. 
The energy difference between \fr{gsH} and the regular solutions
discussed above can be calculated for large $L$ using standard methods
\cite{FSspectrum} and is found to be negative. Considering
other non-regular Bethe Ansatz states in an analogous way, we find that
(\ref{gsH}) is in fact the ground state.

{\it Bosonization}. The low-energy physics of the Hubbard model is
described by a spin-charge separated two-component Luttinger liquid
Hamiltonian \cite{affleck}
\be
H=\sum_{{\tt a}=c,s}\frac{v_{\tt a}}{2}\int dx
\left[(\partial_x\Phi_{\tt a})^2+
(\partial_x\Theta_{\tt a})^2\right],
\label{Hboson}
\ee  
where $v_{c,s}$ are the velocities of the collective charge and spin
degrees of freedom. For $L=0\ {\rm mod}\ 8$ the mode expansions of
the canonical Bose fields
$\Phi_{\tt a}=\varphi_{\tt a}+\bar{\varphi}_{\tt a}$, and their
dual fields $\Theta_{\tt a}=\varphi_{\tt a}-\bar{\varphi}_{\tt a}$ follow
from 
%\begin{widetext}
%\bea
%\varphi_{\tt a}(x,t)\!&=&\!P_{\tt a}+\frac{x-vt}{La_0}Q_{\tt a}
%+\sum_{n=1}^\infty\frac{e^{i\frac{2\pi n}{La_0}(x-vt)}a_{{\tt a},n}+{\rm h.c.}}%{\sqrt{4\pi n}},\nn
%\bar{\varphi}_{\tt a}(x,t)\!&=&\!\bar{P}_{{\tt a}}+\frac{x+vt}{La_0}
%\bar{Q}_{\tt a}+\sum_{n=1}^\infty\frac{e^{-i\frac{2\pi n}{La_0}(x+vt)}\bar{a}_{%{\tt a},n}+{\rm h.c.}
%}{\sqrt{4\pi n}},\nn
%\label{modes}
%\eea
\bea
\bar{\varphi}_{\tt a}(x,t)\!&=&\!\bar{P}_{{\tt a}}+\frac{x_+}{La_0}
\bar{Q}_{\tt a}+\sum_{n=1}^\infty\frac{e^{-i\frac{2\pi
      n}{La_0}x_+}\bar{a}_{{\tt a},n}+{\rm h.c.} }{\sqrt{4\pi n}},\nn
\varphi_{\tt a}(x,t)\!&=&\!P_{\tt a}+\frac{x_-}{La_0}Q_{\tt a}
+\sum_{n=1}^\infty\frac{e^{i\frac{2\pi n}{La_0}x_-}a_{{\tt a},n}+{\rm
    h.c.}}{\sqrt{4\pi n}},
\label{modes}
\eea
where $x_\pm=x\pm vt$ and $a_0$ is the lattice spacing. The
structure of the ground state for $L=0{\ \rm mod\ }8$
is encoded in the zero modes, which have commutations 
relations $[P_{{\tt a}},Q_{\tt a}]=-\frac{i}{2}=-[\bar{P}_{{\tt
      a}},\bar{Q}_{\tt a}]$. The eigenvalues of $Q_{\tt a}$ are 
\bea
q_c&=&\sqrt{\frac{\pi}{8K_c}}\sum_{\sigma=\up,\down}(K_c+1)
m_\sigma+(1-K_c)\bar{m}_\sigma \ ,\nn
q_s&=&\sqrt{\frac{\pi}{2}}\big(m_\up-m_\down)\ ,
\label{zeromodes}
\eea
where $K_c$ is the Luttinger parameter in the charge sector,
$m_\sigma$ are half odd-integer numbers, and the eigenvalues of
$\bar{Q}_{\tt a}$ are obtained by interchanging
$m_\sigma\leftrightarrow\bar{m}_\sigma$. The Hamiltonian then has the
mode expansion 
\be
H=\sum_{{\tt a}=c,s}\frac{v_{\tt a}}{La_0}\Big[Q_{\tt
    a}^2+\bar{Q}_{\tt a}^2+\sum_{n=1}^\infty 2\pi n(
a^\dagger_{{\tt a},n}a_{{\tt a},n}+\bar{a}^\dagger_{{\tt a},n}\bar{a}_{{\tt a},n})
\Big].
\label{modeex}
\ee
There are two degenerate ground states
\be
|\pm\rangle=\frac{1}{\sqrt{2}}\left[|1,0;0,1\rangle\ \pm
  |0,1;1,0\rangle
\right],
\ee
where we have introduced a notation
$|m_\up,m_\down;\bar{m}_\up,\bar{m}_\down\rangle$ for states that are
annihilated by all $a_{{\tt a},n}$, $\bar{a}_{{\tt a},n}$ and have
eigenvalues $q_{\tt a}(m_\sigma,\bar{m}_\sigma)$ and $\bar{q}_{\tt
  a}(m_\sigma,\bar{m}_\sigma)$ of the zero mode operators $Q_{\tt a}$
and $\bar{Q}_{\tt a}$ respectively. In the Hubbard model the
degeneracy between $|+\rangle$ and $|-\rangle$ is removed by the
presence of a marginally irrelevant interaction of spin currents and
the ground state in fact corresponds to $|+\rangle$. Carrying out the
usual conformal map from the cylinder to the plane, we can express
this state in radial quantization as \cite{diF}
\be
|+\rangle\propto\lim_{z,\bar{z}\to 0}
\cos\big(\sqrt{2\pi}\Phi_s(z,\bar{z})\big)|0\rangle\ ,
\label{EXC}
\ee
where $|0\rangle$ is the vacuum state of the free boson theory
\fr{Hboson} on the plane and $z=\exp\big(\frac{2\pi}{La_0}(vt-ix)\big)$,
$\bar{z}=\exp(\frac{2\pi}{La_0}(vt+ix))$.
The key result of the above considerations is that in the Luttinger
liquid approximation to the Hubbard model, the ground state for
$L=0\ {\rm mod}\ 8$ is given by the excited state \fr{EXC} of the free
boson theory on the plane. We note that the operator
$\cos\big(\sqrt{2\pi}\Phi_s\big)$ is not a local operator on the cylinder.

{\it CFT approach to the R\'enyi entropies}. The above considerations
permit us to reduce the calculation of the entanglement entropies in the 
ground state of the Hubbard model to a CFT calculation of the
entanglement in a low-lying excited state. A general approach to the
latter problem has been recently developed by Alcaraz et
al. \cite{abs-11,abs-12} and their main result can be summarized as
follows. The n'th R\'enyi entropy for an excited state of the form
${\cal   O}(0,0)|0\rangle$ is given by
\bea
S_n&=&\frac{c}{6}\big(1+\frac{1}{n}\big)\ln\left[\frac{L}{\pi}
\sin\left(\frac{\pi\ell}{L}\right)\right]+c'_n\nn
&&+\frac{1}{1-n}\ln\left[F_n(\ell/L)\right]+o(L),
\label{SnCFT}
\eea
where $c'_n$ is a ${\cal O}$-independent constant, and the scaling
functions $F_n^{\cal O}(x)$ are given by
\be
F_{n}(x)=\frac{\langle\prod_{k=0}^{n-1}{\cal O}\big(\frac{\pi}{n}(x+2k)\big)
{\cal O}^\dagger\big(\frac{\pi}{n}(-x+2k)\big)\rangle}
{n^{2n(h+\bar{h})}\langle{\cal O}\big(\pi x\big){\cal
    O}^\dagger\big(-\pi x\big)\rangle^n}\ . 
\label{FnO}
\ee
Here $h$ and $\bar{h}$ are the conformal dimensions of the operator
${\cal O}$.
In our case ${\cal O}(x)= 2\cos (\sqrt{2\pi} \Phi_s(x))$ and we need
to evaluate (${\bf\sigma}=\sum_{l=1}^n\sigma_l$)
\be
\langle\prod_{j=1}^{2n}{\cal O}(x_j)\rangle=\!\!
\sum_{\sigma_1,\ldots,\sigma_n=\pm}
\delta_{{\bf \sigma},0}\prod_{i<j}\Big
|2\sin\big(\frac{x_i-x_j}{2}\big)\Big|^{-\sigma_i\sigma_j} ,
\label{Cgas}
\ee
where the $x_j$'s are given by \fr{FnO}. We find that $F_n^s(x)$ can
be expressed as the square root of a determinant, which, surprisingly,
is identical to Eq. (56) of Ref.~\cite{abs-12}. We have succeeded in
expressing this determinant in a form amenable for analytic
continuation in $n$
\bea
\left[F_n(x)\right]^2&=& \prod_{p=1}^n\Big[1-\frac{(n-2 p+1)^2}{n^2}
\sin ^2(\pi  x)\Big] \nn
&=&  \bigg[\Big[ \frac{2\sin (\pi  x)}{n}\Big]^{n}
\frac{ \Gamma \left(\frac{1+n +n \csc (\pi  x) }2 \right)}{
\Gamma \left(\frac{1-n +n \csc (\pi  x) }2 \right)}\bigg]^2.
\label{Fnx}
\eea
Using that for the Hubbard model $c=2$ we then obtain a CFT prediction
for the shell filling effect by combining equations
\fr{Fnx} and \fr{SnCFT}. In order to obtain an expression for the von
Neumann entropy we need to take the limit $n\to 1$, which gives
\be 
S_1=\frac{c}3 \ln \Big(\frac{L}\pi \sin\frac{\pi \ell}L\Big)
+c'_1-F'_1(x)+o(L)\ ,
\label{cfts1}
\ee 
where
\be
F_1'(x)= \ln\big|2 \sin(\pi x)\big| + \psi\bigg(\frac{1}{2\sin(\pi x)}\bigg) + \sin(\pi x).
\ee
Here $\psi(x)$ is the digamma function. We note that both \fr{Fnx} and
\fr{cfts1} apply also to certain excited states in spin chains \cite{abs-12}.
For small $x$, we have $(F_1(x))'= \pi^2 x^2/3+O(x^4)$ in agreement
with the general result in \cite{abs-11}. 

\begin{figure}[ht]
\begin{center}
\includegraphics[width=0.95\linewidth]{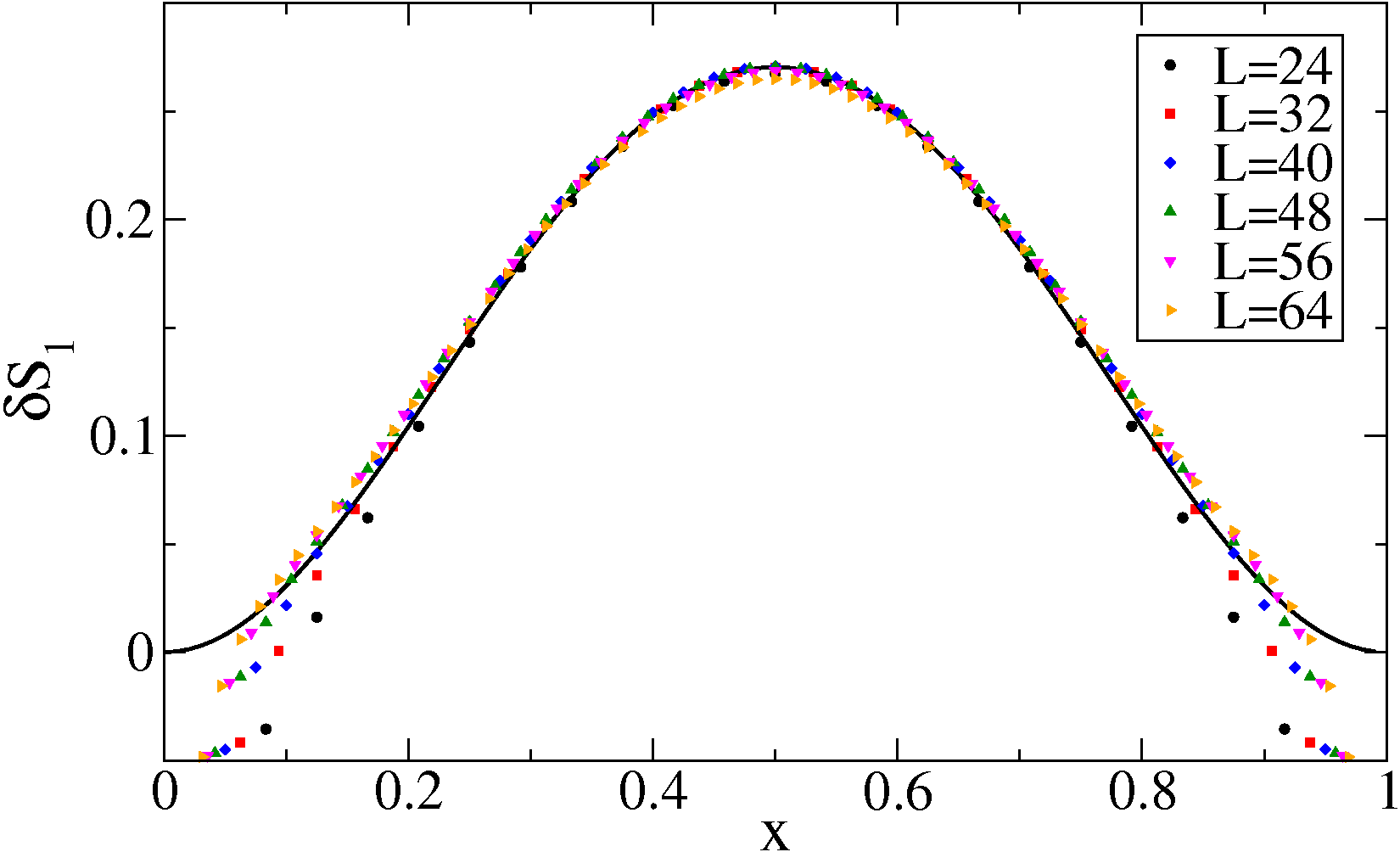}
\end{center}
\caption{ $\delta S_1\equiv S_1-\frac{2}{3}\ln\left[\frac{L}{\pi}
\sin\left(\frac{\pi\ell}{L}\right)\right]-c'_1$ as a function of
  $x=\ell/L$ for $U=0.3t$ and $L=24,32,40,48,56,64$. 
 The constant $c_1'=1.205$ has been adjusted by hand. 
 The solid curve is  $-F_1'(x)$. } 
\label{fig:vN1a}
\end{figure}

{\it Comparison with numerical results}.
We performed extensive DMRG~\cite{DMRG} computations of the periodic
quarter-filled Hubbard model by keeping $M=3000$ states in order to achieve 
satisfactory convergence for periodic systems up to length $L=64$. For small 
values of $U\alt t$ we find good agreement for both $S_1$ and $S_2$ with the 
predictions \fr{cfts1} and \fr{SnCFT}. A representative example is shown in
Fig.~\ref{fig:vN1a}. 
As expected the agreement with the CFT prediction is best for large
block lengths $\ell\sim L/2$ and becomes poor for small $\ell$, when
lattice effects become important. In this region $S_2$ furthermore
exhibits strong oscillatory behaviour as expected \cite{oscillations}.
For larger values of $U\agt t$ the agreement with the CFT predictions
for both $S_1$ and $S_2$ becomes increasingly poor. We now turn to the
origin of these discrepancies. 

{\it Effects of the marginal perturbation}.
It is well know that in the Hubbard model the low-energy Luttinger
liquid Hamiltonian \fr{Hboson} is perturbed by a marginally
irrelevant operator in the spin sector \cite{affleck}. This leads to
logarithmic corrections \cite{logs}, which can be quite important for
small system sizes. The effects of a marginal perturbation on the
ground state entanglement in CFTs was studied in
\cite{cc-10}. These corrections are small for the isotropic Heisenberg
chain \cite{chico} as well as the Hubbard model for $L=4\ {\rm
  mod}\ 8$. However, the effects of the marginal perturbation on the
shell-filling effect are quite large already for moderate values of
$U\agt 2t$. In order to quantify them, we have considered an
extended Hubbard model
\bea
H_{\rm ext}&=&H_{\rm Hubb}
+V_2\sum_{j,\sigma,\sigma'} n_{j,\sigma}n_{j+2,\sigma'}.
\label{HHubbV}
\eea
At weak coupling the main effect of $V_2$ is to reduce the bare
coupling constant of the marginal perturbation. We note that a
nearest-neighbour density-density interaction would be ineffective at
quarter filling and weak coupling \cite{suzu}. We find that increasing
$V_2$ from zero leads to a significant improvement in the agreement
between the CFT prediction \fr{cfts1}, \fr{SnCFT} for the
shell-filling effect for the available system sizes $L\leq 64$.

{\it Conclusions.} We have described a novel shell-filling effect in
entanglement entropies of the 1D quarter-filled Hubbard
model with periodic boundary conditions. We have developed a CFT
approach to calculate the additional contribution to the
R\'enyi entropies, and found good agreement with numerical
computations. The effect, while somewhat unexpected, has a simple
origin: for certain ratios of lattice lengths to particle numbers in
multi-component systems, the ground state cannot be thought of in
terms of a product of Fermi seas (in general these will consist of
appropriate elementary excitations), but is in fact a linear
combination of different such seas. This suggests similarities
with the results obtained \cite{grover,ben} for the entanglement of linear
combinations of degenerate ground states. However, in our case the
ground state is unique for $U>0$ (and fixed $S^z=0$) and is thus not 
based on a degeneracy.
Finally, we expect shell-filling effects to exist for interacting
bosons as well as for fermions in one dimension, as well as in
higher dimensional critical systems. They can for example play a
role in numerical studies of two-dimensional gapless spin liquids,
which display a spinon Fermi surface~\cite{LeeLee,Motrunich,Yang}.

{\it Acknowledgments}. We are grateful to F. Alcaraz, I. Affleck, J. Cardy,
M. Fagotti and N. Robinson for helpful discussions.  
This work was supported by the EPSRC under grants EP/I032487/1 and
EP/J014885/1 (FHLE),  the ERC under  Starting Grant
279391 EDEQS (PC) and the National Science Foundation under grant NSF
PHY11-25915 (FHLE, AML and PC). We thank the GGI in Florence and the KITP
in Santa Barbara for hospitality.

\end{document}